\documentclass[onecolumn,amsmath,amssymb,amsfonts,aps,superscriptaddress]{revtex4-2}

\usepackage{graphicx}
\graphicspath{{figures/}}
\usepackage{epstopdf}
\usepackage{gensymb}
\usepackage{xcolor}
\usepackage{hyperref}
\hypersetup{%
        colorlinks,
        breaklinks=true,
        plainpages=false,%
        citecolor=blue,
        linkcolor=blue,
        urlcolor=blue,
        bookmarksopen=true,%
        bookmarksnumbered=false,%
        bookmarksdepth=5%
}

\begin{document}

\title[Solid-state \textsuperscript{3}He NMR of superconducting Rb\textsubscript{3}(\textsuperscript{3}He@C\textsubscript{60})]{Solid-state \texorpdfstring{\textsuperscript{3}He}{3He} NMR of the superconducting rubidium endofulleride \texorpdfstring{Rb\textsubscript{3}(\textsuperscript{3}He@C\textsubscript{60})}{Rb3(3He@C60)}.}

\author{\firstname{Murari} \surname{Soundararajan}}\email{ms3v21@soton.ac.uk}
\author{\firstname{George R.} \surname{Bacanu}}
\author{\firstname{Francesco} \surname{Giustiniano}}
\author{\firstname{Mark C} \surname{Walkey}}
\author{\firstname{Gabriela} \surname{Hoffman}}
\author{\firstname{Marina} \surname{Carravetta}}
\affiliation{Department of Chemistry, University of Southampton, Southampton SO17 1BJ, UK}
\author{\firstname{Martin R.} \surname{Lees}}
\affiliation{Physics Department, University of Warwick, Coventry CV4 7AL, UK}
\author{\firstname{Richard J.} \surname{Whitby}}
\author{\firstname{Malcolm H.} \surname{Levitt}}
\affiliation{Department of Chemistry, University of Southampton, Southampton SO17 1BJ, UK}

\date{23 August, 2023}

\begin{abstract}
    A new variant of the superconducting fulleride Rb\textsubscript{3}C\textsubscript{60} is presented, with \textsuperscript{3}He atoms encapsulated in the C\textsubscript{60} cages. The \textsuperscript{3}He nuclei act as sensitive NMR probes embedded in the material. The superconducting and normal states are characterised by \textsuperscript{3}He NMR. Evidence is found for co-existing vortex liquid and vortex solid phases below the superconducting transition temperature. A strong dependence of the spin-lattice relaxation time constant on spectral frequency is observed in the superconducting state, as revealed by two-dimensional NMR utilising an inverse Laplace transform. Surprisingly, this phenomenon persists, in attenuated form, at temperatures well above the superconducting transition.
\end{abstract}

\maketitle

\section{Introduction}\label{intro}
Reactions of fullerene $\mathrm{C_{60}}$ with alkali metals give rise to a wide range of salts called \emph{fullerides}, which consist of negatively-charged $\mathrm{C_{60}^{n-}}$ anions ($n\in\{1\ldots6\}$) separated by positive alkali metal cations~\cite{holczer_AlkaliFullerideSuperconductors_1991}.  
Superconductivity in the alkali metal fullerides was first observed for $\mathrm{K_3C_{60}}$, which has a superconducting transition temperature ($T_C$) of 18~K \cite{hebard_Superconductivity18_1991}. The transition temperature $T_C$ increases as the size of the alkali metal cation (or unit cell volume) increases, reaching 33~K in $\mathrm{Cs_2RbC_{60}}$ at ambient pressure~\cite{gunnarsson_Superconductivityfullerides_1997}. The rubidium fulleride $\mathrm{Rb_3C_{60}}$, which is the topic of this paper, has a superconducting transition temperature of around 30~K~\cite{rosseinsky_Superconductivity28_1991}.

Despite their relatively high transition temperatures $T_C$ and upper critical fields ($H_{C2}$), alkali metal fullerides have generally been understood as s-wave superconductors following conventional Bardeen-Cooper-Schrieffer (BCS) theory~\cite{pennington_Nuclearmagnetic_1996}. More recently, phase diagrams of Cs-containing fullerides as well as other large volume fullerides over wide ranges of temperature and pressure have been reported, displaying unconventional metallic and superconducting states \cite{zadik_Optimizedunconventional_2015,takabayashi_DisorderFreeNonBCS_2009}. These reports suggest an important role played by strong electron correlations while retaining s-wave pairing, placing fullerides in a category distinct from both conventional BCS superconductors and high-temperature superconductors such as cuprates \cite{takabayashi_Unconventionalhigh_2016}.

NMR is a valuable technique for studying superconductors both in their normal and superconducting states. The shape and position of the NMR spectrum and nuclear spin-lattice relaxation time ($T_1$) provide information about electronic interactions and  magnetic field distributions inside the material, and also as direct probes of superconducting parameters such as the band gap~\cite{maclaughlin_MagneticResonance_1976, f.smith_StudyMechanisms_2006, mounce_Nuclearmagnetic_2011}. Rubidium fulleride $\mathrm{Rb_3C_{60}}$ has been extensively studied by both $\mathrm{^{13}C}$ and $\mathrm{^{87}Rb}$ NMR~\cite{zimmer_Analysis87_1993}. 
However, the interpretation of the NMR data in terms of the electronic structure and dynamics is made more complicated by the large chemical shift anisotropy of the fullerene $\mathrm{^{13}C}$ nuclei, and by the electric quadrupole interactions of the Rb nuclides.

Each molecule of fullerene $\mathrm{C_{60}}$ is a symmetrical carbon cage enclosing a central cavity with an internal diameter of $\sim3.4$~\AA. Supramolecular complexes known as \emph{endofullerenes} may be formed, in which each fullerene cavity accommodates a single atom or molecule. These complexes are 
denoted A@$\mathrm{C_{60}}$, where A is the endohedral species. Noble gas endofullerenes such as He@$\mathrm{C_{60}}$ and Xe@$\mathrm{C_{60}}$ were first synthesized in very low yield by high-temperature methods~\cite{saunders_Incorporationhelium_1994}. The remarkable endofullerene N@$\mathrm{C_{60}}$ has been produced by ion bombardment of $\mathrm{C_{60}}$~\cite{almeidamurphy_ObservationAtomlike_1996}. Recently, the multistep synthetic procedure known as ``molecular surgery" has made it possible to synthesize, in high yield and high purity, a wide range of atomic and molecular endofullerenes, incorporating species such as $\mathrm{H_2}$, $\mathrm{HD}$, $\mathrm{H_2O}$, $\mathrm{HF}$, $\mathrm{CH_4}$, $\mathrm{He}$, $\mathrm{Ne}$, $\mathrm{Ar}$ and $\mathrm{Kr}$~\cite{bloodworth_Synthesisendohedral_2022}. These systems have been extensively studied by NMR, infrared spectroscopy, terahertz spectroscopy, and inelastic neutron scattering~\cite{aouane_combinedinelastic_2023,bacanu_Finestructure_2020,bloodworth_CH4@C60_2019,hoffman_Synthesis83Kr_2022,jafari_NeAr_2023,jafari_Terahertzspectroscopy_2022,krachmalnicoff_HF@C60_2016,shugai_Infraredspectroscopy_2021}. These studies have revealed the spatial quantisation of the confined atoms or molecules, the rich interactions of the quantised translational modes with the molecular rotations and vibrations, and the spin isomerism of confined symmetrical species such as $\mathrm{H_2}$ and $\mathrm{H_2O}$~\cite{turro_DemonstrationChemical_2008, meier_Electricaldetection_2015}. ``Non-bonded" internuclear J-couplings have been observed between the nucleus of endohedral $\mathrm{^3He}$ and the $\mathrm{^{13}C}$ nuclei of the encapsulating cage in $\mathrm{^3He@C_{60}}$~\cite{bacanu_InternuclearCoupling_2020}. The potential energy function describing the interactions of endohedral He atoms with the fullerene cages was elucidated by terahertz spectroscopy and neutron scattering~\cite{bacanu_Experimentaldetermination_2021}.

The availability of endofullerenes by molecular surgery suggests the possibility of creating \emph{endofullerides} through the reaction of an endofullerene with an alkali metal, generating materials in which endofulleride anions such as $\mathrm{A@C_{60}^{3-}}$ are separated by alkali metal cations. Endofulleride salts such as $\mathrm{Rb_3(H_2@C_{60})}$ and $\mathrm{Rb_3(H_2O@C_{60})}$ have been synthesised and studied by solid-state NMR, both in the metallic and superconducting states~\cite{bounds_Nuclearmagnetic_2016}.

Endofullerides with A=$\mathrm{^3He}$ are particularly attractive for NMR investigations, since the $\mathrm{^3He}$ nucleus is spin-1/2, has a large gyromagnetic ratio, no quadrupole moment, and should lack chemical shift anisotropy when located at the centre of a symmetrical $\mathrm{C_{60}}$ cage. The zero natural abundance of $\mathrm{^3He}$ ensures the absence of background signals from the sample container, probe parts, etc. which can be a significant nuisance factor for other nuclides~\cite{saunders_Probinginterior_1994}. In this paper we describe some preliminary solid-state $\mathrm{^3He}$ NMR observations of the He endofulleride $\mathrm{Rb_3(^3He@C_{60})}$, over a range of temperatures spanning the high-temperature metallic phase and the low-temperature superconducting phase. We show that the endohedral $\mathrm{^3He}$ atoms have a negligible influence on the superconducting properties of $\mathrm{Rb_3C_{60}}$ and obtain some novel insights into superconductivity in this material using relaxation and spectral information from $\mathrm{^3He}$ NMR. The endohedral spin-1/2 $\mathrm{^3He}$ nuclei serve as excellent probes of the electronic structure and dynamics, and the internal magnetic fields of the material, with high NMR sensitivity and zero background interference.

\section{Experimental Details}\label{sec:exp}
\subsection{Samples}\label{subsec:samples}

The endofullerene $\mathrm{^3He@C_{60}}$ used in the synthesis of the endofulleride $\mathrm{Rb_3(^3He@C_{60})}$ was prepared as in \cite{hoffman_SolidState_2021}. $\mathrm{Rb_3(^3He@C_{60})}$ with $22\%$ of the $\mathrm{C_{60}}$ containing a $\mathrm{^3He}$ atom (a $22\%$ ``filling factor'') was prepared by a refinement of the method reported in \cite{mccauley_Synthesisstructure_1991} for $\mathrm{Rb_3C_{60}}$. $\mathrm{Rb_6C_{60}}$ was prepared as a pure (by X-ray diffraction) free-flowing powder by reacting sublimed $\mathrm{C_{60}}$ with excess Rb metal (10 equivalents, $400\ ^\circ\mathrm{C}$, 2 days) in a glass tube sealed under vacuum, then removing the excess Rb by treatment in a thermal gradient (tube furnace at $400\ ^\circ\mathrm{C}$, 3 days, condensing end of the tube 3~cm out of the furnace). In a glove box the $\mathrm{Rb_6C_{60}}$ was weighed, then the calculated stoichiometric amount of $\mathrm{He@C_{60}}$ ($44\%$ filled) needed to produce the ternary salt in its pure form added and well mixed. After sealing in a glass tube the mixture was heated for 4 days at $450\ ^\circ\mathrm{C}$ affording pure (by X-ray diffraction, see Figure S.3) $\mathrm{Rb_3(^3He@C_{60})}$ with a filling factor of $22\%$. Samples used for XRD were taken from the same synthesis batch as the NMR sample. No amorphous signatures or spurious crystalline peaks are seen in the XRD, indicating that impurities are  limited to 5-10~\% of the sample at most. Reference $\mathrm{^{87}Rb}$ NMR spectra of the sample at room temperature (see section 3 of the SI) display only the expected $\mathrm{Rb_3(^3He@C_{60})}$ peaks, further indicating that no significant quantity of impurities are present.

40~mg of $\mathrm{Rb_3(^3He@C_{60})}$ was filled into a 4~mm outer diameter, 2~mm inner diameter borosilicate glass tube inside a glove box kept under an inert nitrogen environment. The tube was filled with 0.25~bar of helium gas ($\mathrm{^4He}$ to ensure good thermal contact with the environment) and sealed with a flame. The same sample was used for NMR measurements and for magnetic susceptibility measurements.

\begin{figure}[ht]
    \centering
    \includegraphics[width=0.9\textwidth]{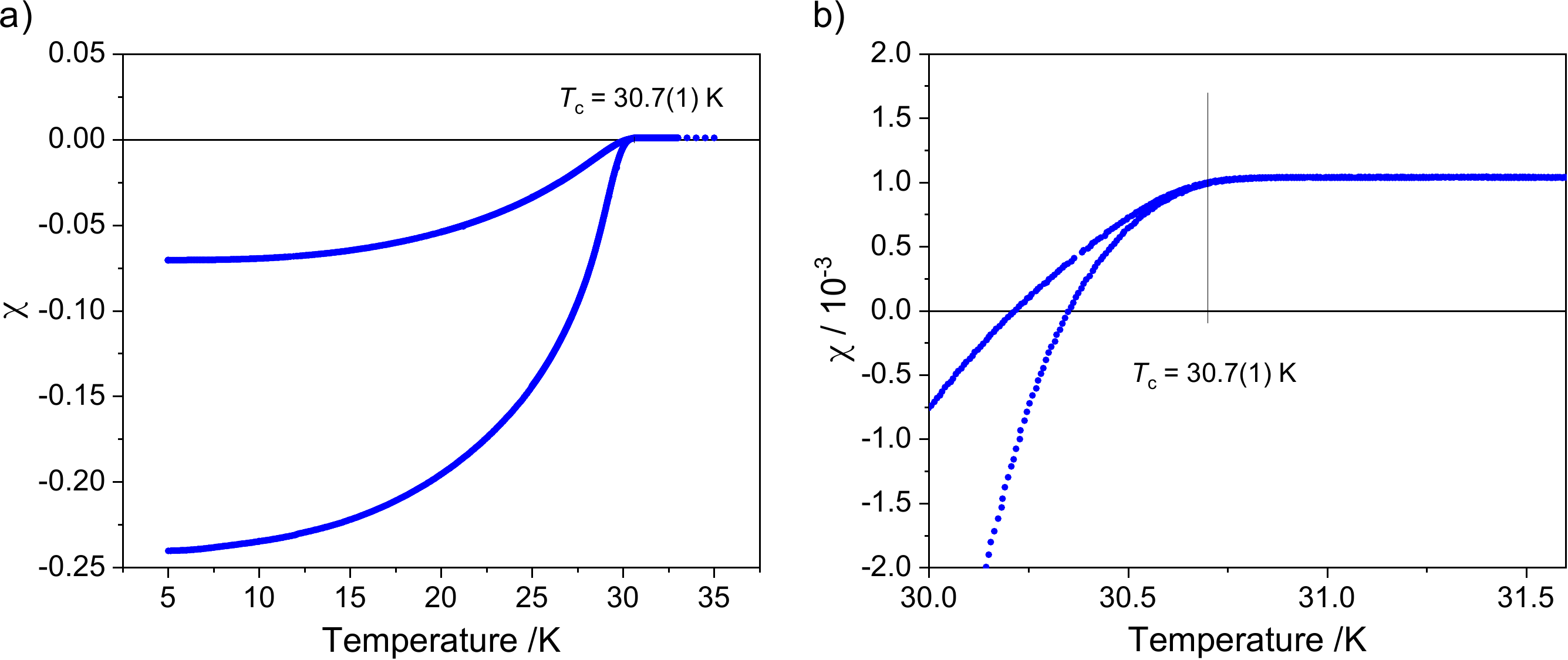}
    \caption{DC susceptibility ($\chi$) of $\mathrm{Rb_3(^3He@C_{60})}$ at an externally applied magnetic field $\mu_0H\,=2.5\mathrm{\,mT}$.
    Field cooled (upper trace) and zero field cooled (lower trace) measurements are plotted on the same graph. a) $\chi$ versus temperature over the entire temperature sweep range of 5~K to 35~K. b) $\chi$ versus temperature near 30~K from which a $T_C$ of ($30.7\pm0.1$)~K can be estimated.
    }
    
    \label{fig:susceptibility}
\end{figure}

\subsection{Instrumentation}\label{subsec:instr}
\subsubsection{Helium-3 NMR}
NMR experiments were performed at 14.1~T on a Bruker AS 600 WB magnet equipped with an AVANCE NEO console and fitted with an Oxford Instruments Spectrostat flow cryostat. A cryogenic $\mathrm{^3He}$/X probe designed by NMR Service GmbH and tuned to the $\mathrm{^3He}$ Larmor frequency of 457.401426 MHz was used. A detailed description of the construction and NMR performance of the probe is given in section 1 of the SI.
The experiments used a $\mathrm{^3He}$ radiofrequency amplitude corresponding to a nutation frequency of  $\sim$32 kHz at ambient temperature. A nutation frequency of $\sim$26 kHz was used below 35~K to prevent probe arcing.

\subsubsection{Magnetic susceptibility}
A Quantum Design (QD) Magnetic Property Measurement System (MPMS) SQuID magnetometer was used to collect DC magnetic susceptibility versus temperature data between 5 and 300~K in zero-field-cooled warming and field-cooled cooling modes in an applied magnetic field $\mu_0H$ of 2.5~mT.

\section{Results and Discussion}\label{sec:results}
\subsection{Susceptibility measurements}
In order to probe the superconducting transition of $\mathrm{Rb_3(^3He@C_{60})}$, the magnetic susceptibility of the sample was measured as a function of temperature at an external applied magnetic field $\mu_0H$ of 2.5~mT, well below the lower critical field $\mu_0H_{C1}=11.4\mathrm{\,mT}$ of $\mathrm{Rb_3C_{60}}$~\cite{politis_PENETRATIONDEPTH_1992}.

Field cooled (FC) and zero field cooled (ZFC) measurements of the DC susceptibility of the $\mathrm{Rb_3(^3He@C_{60})}$ sample are presented in \hyperref[{fig:susceptibility}]{Figure~\ref*{fig:susceptibility}} in the temperature range of 5 to 35~K. The data indicates a superconducting transition at $T_C = (30.7\pm0.1)$~K. This is consistent with the $T_C$ estimates of $\mathrm{Rb_3C_{60}}$ reported in the literature, which lie between 25 and 31~K~\cite{politis_PENETRATIONDEPTH_1992, rosseinsky_Superconductivity28_1991, louis_SuperconductingProperties_1994}. Hence these data strongly indicate that the presence of $\mathrm{^3He}$ inside the fullerene cages does not significantly influence the onset of superconductivity.

The shielding fraction extracted from the ZFC curve is about 27\% and the Meissner fraction extracted from the FC curve is about 7\%. Low superconducting fractions such as these are commonly observed in fulleride superconductors. Politis \textit{et al. } \cite{politis_PENETRATIONDEPTH_1992} analyse the particle size distribution in their samples of $\mathrm{Rb_3C_{60}}$ and conclude that a significant proportion of particles are of comparable or smaller size than the magnetic field penetration depth at 0 K. They state that the shielding fraction, even at low applied magnetic fields, is an underestimate of the superconducting fraction. The large discrepancy between Meissner and shielding fractions is a known effect in type-II superconductors due to flux pinning in the mixed state just below $T_C$ \cite{dahlke_superconductivity_2000} and has been analysed in detail for high-temperature superconductors like cuprates \cite{tomioka_meissner_1994}. From the spectroscopic and diffraction data, which gives no evidence of significant impurities, and the documented difficulties in the interpretation of observed Meissner fractions for finely divided powders, we contend that a large fraction of the sample is indeed in the superconducting phase of $\mathrm{Rb_3(^3He@C_{60})}$ below 30~K, despite the low observed Meissner fraction.

\subsection{NMR spectra}

\begin{figure}[ht]
    \centering
    \includegraphics[width=0.9\textwidth]{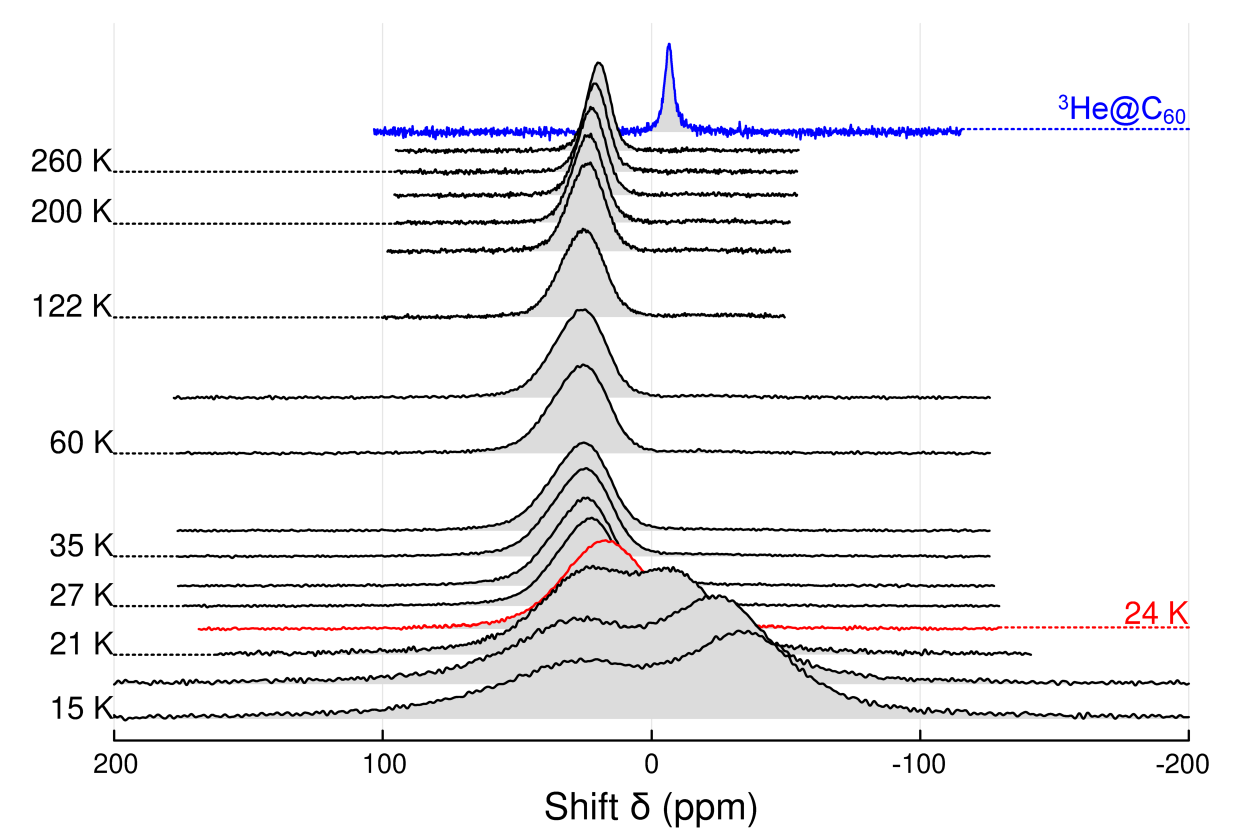}
    \caption{$\mathrm{^3He}$ NMR spectra of $\mathrm{Rb_3(^3He@C_{60})}$ from 15~K to 290~K presented as a waterfall plot along with a reference spectrum of $\mathrm{^3He@C_{60}}$ at 292~K. For clarity, the vertical scale of each spectrum has been adjusted to give a uniform vertical height over the series.
    The superconducting transition is seen as a shift in the mean peak position of the spectrum at 24~K, plotted in red. The split peak below $T_C$ is attributed to the coexistence of vortex solid and vortex liquid phases (see text).}
    \label{fig:waterfall}
\end{figure}

$\mathrm{^3He}$ NMR measurements on the endofulleride $\mathrm{Rb_3(^3He@C_{60})}$ were performed at temperatures between 15~K and 290~K.
The series of spectra is shown in \hyperref[{fig:waterfall}]{Figure~\ref*{fig:waterfall}}, which also includes the $\mathrm{^3He}$ NMR spectrum of solid $\mathrm{^3He@C_{60}}$, obtained in a separate measurement at 292~K. 
This signal, set to $\delta = -6.3\mathrm{\,ppm}$ relative to $\mathrm{^3He}$ gas dissolved in 1-methylnaphthalene~\cite{saunders_Probinginterior_1994}, was used as a secondary external reference for the $\mathrm{Rb_3(^3He@C_{60})}$ spectra. 

At room temperature, the $\mathrm{^3He}$ resonance of $\mathrm{Rb_3(^3He@C_{60})}$  is shifted from that of $\mathrm{^3He@C_{60}}$ by +25.9~ppm (i.e. in the direction of less shielding, or higher absolute frequency). This shift is a superposition of a chemical shift due to the local electronic environment of the endohedral $\mathrm{^3He}$ atoms, and a Knight shift due to hyperfine couplings with the conduction electrons in the high-temperature metallic phase. For comparison, the $\mathrm{^{13}C}$ nuclei of $\mathrm{Rb_3C_{60}}$ experience a shift of $+60\mathrm{\,ppm}$~\cite{zimmer_Analysis87_1993}.

NMR spectra below 24~K are broadened and shifted to low frequency (high shielding), relative to those obtained at higher temperatures, as seen in \hyperref[{fig:waterfall}]{Figure~\ref*{fig:waterfall}}. The shift and broadening is attributed to the diamagnetism of the superconducting state and the inhomogeneous magnetic field associated with the Abrisokov vortex lattice~\cite{brandt_Magneticfieldvariance_1991, mounce_Nuclearmagnetic_2011}. The spectra below 21~K display a split peak. The low frequency (more shielded) peak appears at 
$\sim$-5~ppm
at 21~K, and broadens and shifts to lower frequency (higher shielding) as the temperature goes down, reaching  
-34~ppm at 15~K.
A similar effect has been reported in the $^{17}$O NMR of cuprate superconductors and has been interpreted in terms of a vortex liquid state coexisting with a frozen vortex lattice \cite{reyes_Vortexmelting_1997}. The observed increase in the relative intensity of the lower frequency (high shielding, right-hand) peak as the temperature is reduced below $T_C$ is consistent with an increased fraction of the frozen vortex phase. Such vortex phase coexistence in $\mathrm{Rb_3C_{60}}$ has been observed through $\mathrm{^{87}Rb}$ and $\mathrm{^{13}C}$ transverse relaxation ($T_2$) measurements \cite{zimmer_Vortexdynamics_1996} but not as a splitting in the NMR spectrum, likely due to the poor resolution and the large $\mathrm{^{87}Rb}$ and $\mathrm{^{13}C}$ linewidths.

\begin{figure}[ht]
    \centering
    \includegraphics[width=0.9\textwidth]{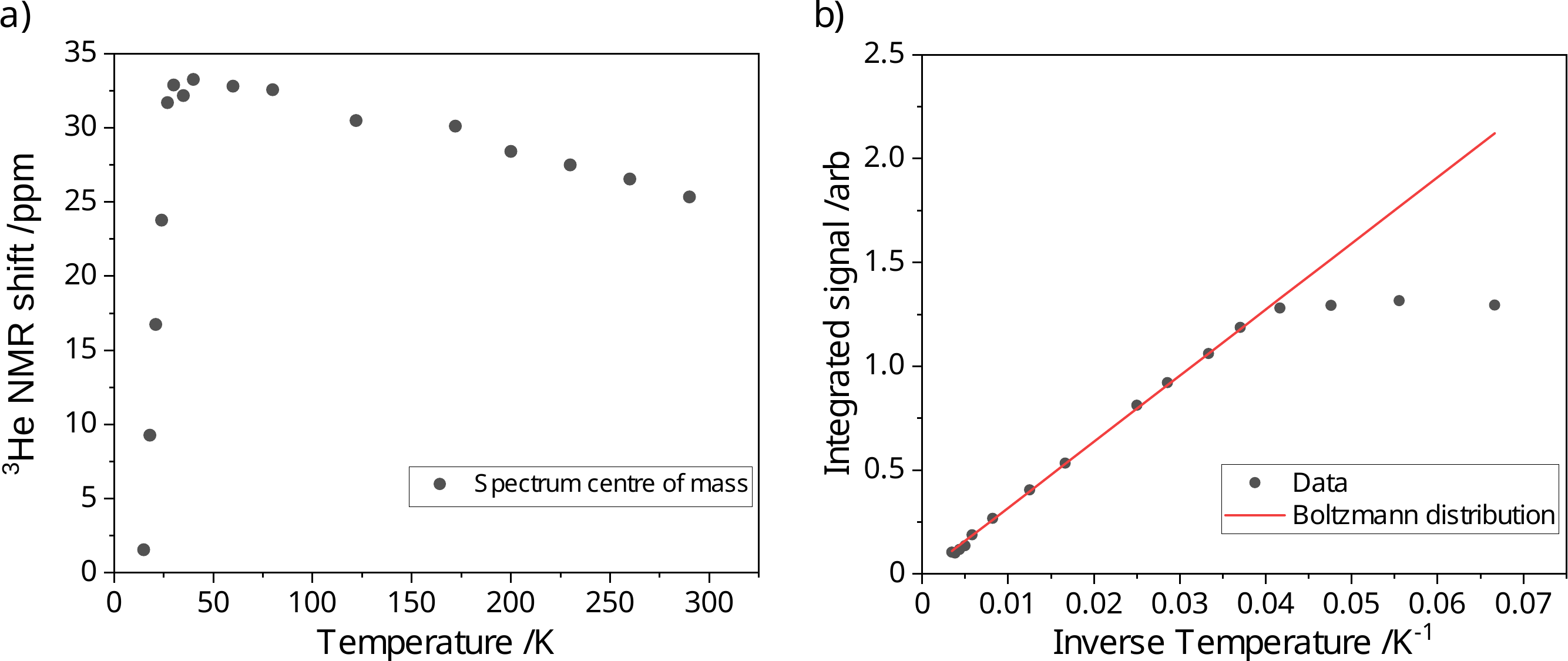}
    \caption{$\mathrm{^3He}$ NMR shifts and intensities of $\mathrm{Rb_3(^3He@C_{60})}$ as a function of temperature. (a)~Estimated shifts as a function of temperature.  The shifts are calculated as the relative position of the centre-of-mass of the spectrum in $\mathrm{Rb_3(^3He@C_{60})}$ relative to that of $\mathrm{^3He@C_{60}}$. (b)~Integrated intensities of the NMR spectra as a function of temperature. The expected Curie law dependence of the $\mathrm{^3He}$ polarisation at 14.1~T is shown in red.}
    \label{fig:shifts}
\end{figure}

The $\mathrm{^3He}$ shift, evaluated as the shift of the centre of mass of the NMR spectrum relative to the $\mathrm{^3He@C_{60}}$ spectrum, is shown for temperatures between 15~K and 290~K in \hyperref[{fig:shifts}]{Figure~\ref*{fig:shifts}(a)}. As the temperature is decreased starting from room temperature, the $\mathrm{^3He}$ shift increases monotonically until the superconducting transition is reached at 24~K. Further cooling causes a dramatic decrease in the $\mathrm{^3He}$ shift.

The temperature-dependence of the $\mathrm{^3He}$ shift above $T_C$ is unexpected. As explained in Section 1 of the SI, this shift cannot be due to instrumental effects. The shift is a superposition of the chemical shift and the Knight shift. The Knight shift is expected to be temperature-independent for metallic conductors within the Pauli approximation~\cite{vanderklink_NMRmetals_2000}, while the $\mathrm{^3He}$ chemical shift is also expected to be roughly temperature-independent. Hence, the total $\mathrm{^3He}$ shift is expected to be almost temperature-independent in the normal metallic state. However, this is not the observation. A similar temperature-dependence of the shift in the metallic state has been observed for $\mathrm{^{13}C}$ and for $\mathrm{^{87}Rb}$ in several alkali metal fullerides~\cite{tycko_Electronicproperties_1992, zimmer_Analysis87_1993}.

The superconducting transition temperature of 24~K is lower than the 30~K indicated by the susceptibility measurements (\hyperref[{fig:susceptibility}]{Figure~\ref*{fig:susceptibility}}). This is presumably due to the much higher magnetic field used in the NMR measurements (14.1~T). The depression of $T_C$ by a strong applied field is a common feature for all superconductors~\cite{helfand_TemperaturePurity_1966}. 

The integrated signal intensity is presented as a function of inverse temperature in \hyperref[{fig:shifts}]{Figure~\ref*{fig:shifts}(b)}, alongside the expected functional form of the $\mathrm{^3He}$ Boltzmann polarisation at 14.1~T, which is proportional to $\tanh{(\hbar\omega_0/k_BT)}$. Over the relevant range of temperature, this corresponds closely to the Curie law (polarisation proportional to $T^{-1}$).
The experimental signal intensities follow the Curie law closely above $T_C\simeq24\mathrm{\,K}$ but deviate sharply below $T_C$ becoming almost temperature-independent at very low temperatures. The reason for this apparent deviation from the Curie law is not known at this point. Since the spectra below $T_C$ were acquired with 4 scans and a fixed recycling delay of 300~s, one possibility is that the increase in $T_1$ with decreasing $T$ prevents the nuclear spin system from reaching complete thermal equilibrium with the lattice at the lowest temperatures, within the timeframe of the experiment. Another possibility is incomplete excitation of the NMR spectrum as the linewidth increases considerably at 21~K and below. A related effect is that as the temperature goes down, an increasing fraction of the sample experiences magnetic fields that are so inhomogeneous that the $\mathrm{^3He}$ nuclei become essentially NMR-silent.

\subsection{Spin-lattice relaxation}\label{subsec:T1}
\begin{figure}[ht]
    \centering
    \includegraphics[width=0.9\textwidth]{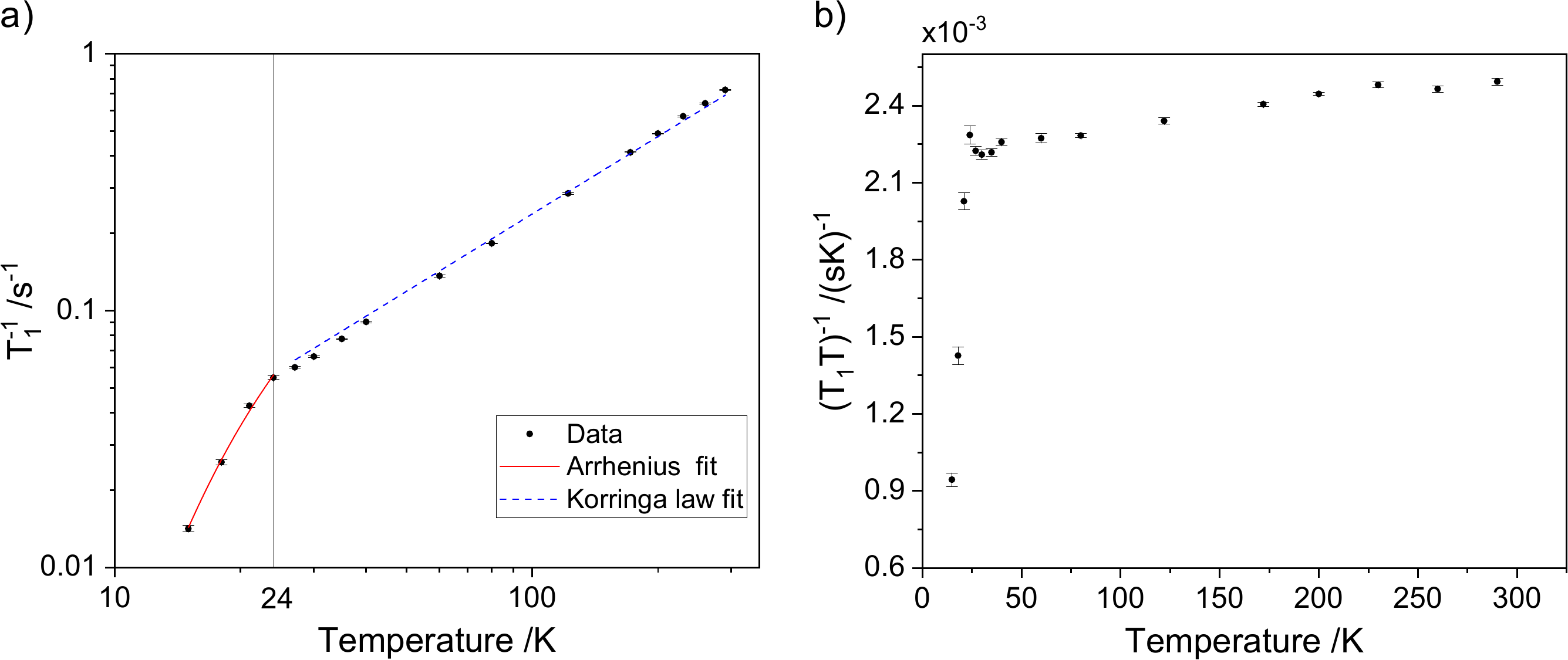}
    \caption{Temperature dependence of the $\mathrm{^3He}$ $T_1$. (a) Black circles: Experimentally determined relaxation rate constants ($T_1^{-1}$) and their error bars. The data points below 24~K are fit to an Arrhenius function (red line), yielding an energy gap $2\Delta$ of $\left(109 \pm 6\right)$~K. The data points above 24~K are fit to a straight line (dashed blue line) corresponding to a best fit~Korringa product of $(T_1T)^{-1} = 0.0024\,\mathrm{(s\,K)^{-1}}$. (b)~Korringa product $(T_1T)^{-1}$ as a function of temperature.}
    \label{fig:R1_v_T}
\end{figure}

\hyperref[{fig:R1_v_T}]{Figure~\ref*{fig:R1_v_T}(a)} shows the results of $T_1$ measurements performed using the inversion-recovery method at temperatures ranging from 15~K to 290~K.
The recovery curves are accurately monoexponential above 21~K. However, significant deviations from single-exponential magnetisation recovery trajectories appear at temperatures of 21~K and below. Furthermore, as discussed later, the $T_1$ values exhibit strong spectral inhomogeneity at low temperatures. 

The $T_1$ values plotted in \hyperref[{fig:R1_v_T}]{Figure~\ref*{fig:R1_v_T}(a)} are simply equal to the monoexponential time constants in the high-temperature region. In the low-temperature region, where the magnetisation recovery is clearly multi-exponential, the data was fit to a biexponential recovery function. In this regime, the plotted values of $T_1$ are an average of the two biexponential time constants, weighted by their coefficients in the biexponential fit.

In the normal state, under the Pauli approximation and in the absence of strong electron correlations, the $T_1$ at a temperature $T$ is related to the~Knight shift $K$ by the~Korringa relation~\cite{vanderklink_NMRmetals_2000}:
\begin{equation}\label{eq:Korringa}
    K^2T_1T = S,
\end{equation}
where the Korringa constant $S$ is given by
\begin{equation*}
    S = \frac{\mu_B^2}{\pi\hbar k_B\gamma_n^2}
\end{equation*}
and $\mu_B$ is the Bohr magneton, $\hbar$ is the reduced Planck constant, $k_B$ is the Boltzmann constant and $\gamma_n$ is the gyromagnetic ratio of the nucleus. For $\mathrm{^3He}$, the value of $S$ is $\sim4.54\times10^{-7}\mathrm{\,s\,K}$. 

If the~Knight shift is independent of temperature, as expected for a normal metal, the~Korringa product $(T_1T)^{-1}$ is temperature-independent. This hypothesis is tested in \hyperref[{fig:R1_v_T}]{Figure~\ref*{fig:R1_v_T}(b)}, which shows the~Korringa product as a function of temperature for the $\mathrm{^3He}$ resonance of $\mathrm{Rb_3(^3He@C_{60})}$. The parameter $(T_1T)^{-1}$ is indeed almost temperature-independent in the normal state, increasing only slightly with increasing temperature, from a value of 0.0022~$\mathrm{(s\,K)^{-1}}$ at 27~K to 0.0025~$\mathrm{(s\,K)^{-1}}$ at 290~K. Similar deviations from the~Korringa law are often observed in fullerides~\cite{tycko_Electronicproperties_1992}. 

In the superconducting state, for a superconducting energy gap $2\Delta$, the relaxation rate can be modelled using an Arrhenius law \cite{masuda_NuclearSpinLattice_1962} as
\begin{equation*}
    T_1^{-1} \propto e^{-\Delta/T}.
\end{equation*}
Fitting the relaxation data below $T_C$ to a function of this form, we obtain an energy gap $2\Delta$ of $(109 \pm 6)$~K. This yields a $2\Delta/T_C$ of $\mathrm{4.5 \pm 0.8}$, which agrees with the value of 4.1 estimated using $\mathrm{^{13}C}$ NMR by Tycko \textit{et. al.}~\cite{tycko_Electronicproperties_1992}.

The peak splitting and multi-exponentiality in the recovery curves below 21~K was explored further by examining the frequency dependence of the $T_1$ across the NMR spectrum at low temperature. The inverse Laplace transform (ILT) is a numerical analysis technique which decomposes an input signal into individual exponential decay components. It is often used in NMR to analyse samples with overlapping spectral features that are unresolvable in the frequency domain on the basis of their $T_1$ and $T_2$ relaxation behaviour~\cite{lee_two-dimensional_1993,moraes_transformada_2021,telkki_ultrafast_2021}. An extension to this approach was proposed in~\cite{lupulescu_relaxation-assisted_2003} in which the ILT is performed pointwise across the NMR spectrum to obtain a so-called relaxation-assisted separation (RAS) NMR spectrum with signal resolution in the frequency and relaxation domains. RAS NMR spectra of our sample at 15~K, 60~K and 290~K, with the noise reduced by averaging over a moving window of 51 points in the frequency domain, are presented as contour maps below. Additional RAS NMR spectra above $T_C$, at 30~K, 35~K and 40~K, are presented in the supplementary information (Figures S.6-S.8).

\begin{figure}[t]
    \centering
    \includegraphics[width=0.9\textwidth]{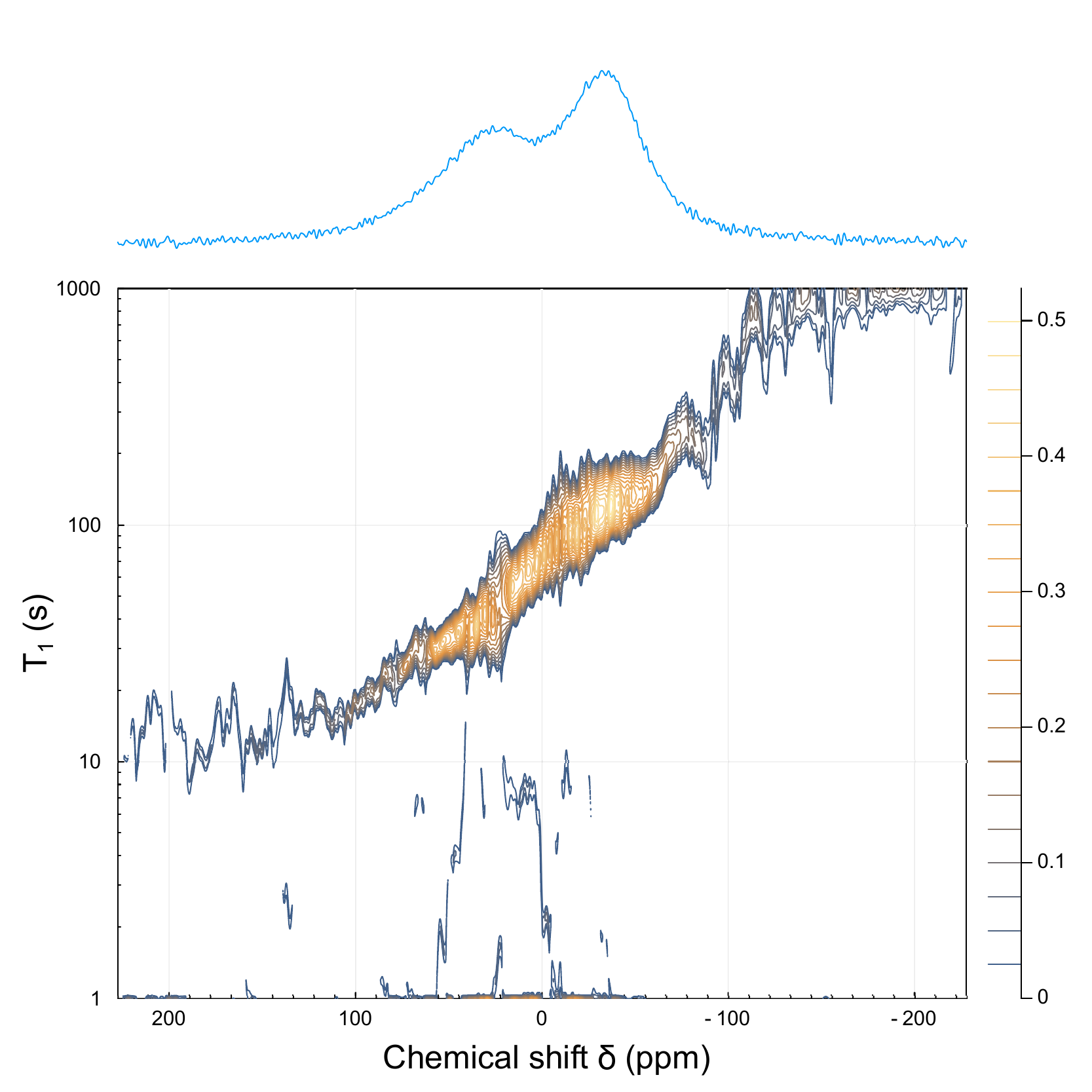}
    \caption{$\mathrm{^3He}$ RAS NMR spectrum~\cite{lupulescu_relaxation-assisted_2003} at 15~K. The image is presented as a contour map. The roughly diagonal ridge indicates a strong, and continuous, correlation between the value of $T_1$ and the $\mathrm{^3He}$ resonance frequency. Note the logarithmic scale used for $T_1$. The 15~K NMR spectrum, extracted from the last slice of the inversion recovery data used as the input dataset, is displayed above the contour map.}
    \label{fig:ilft_15}
\end{figure}

\begin{figure}[ht]
    \centering
    \includegraphics[width=0.9\textwidth]{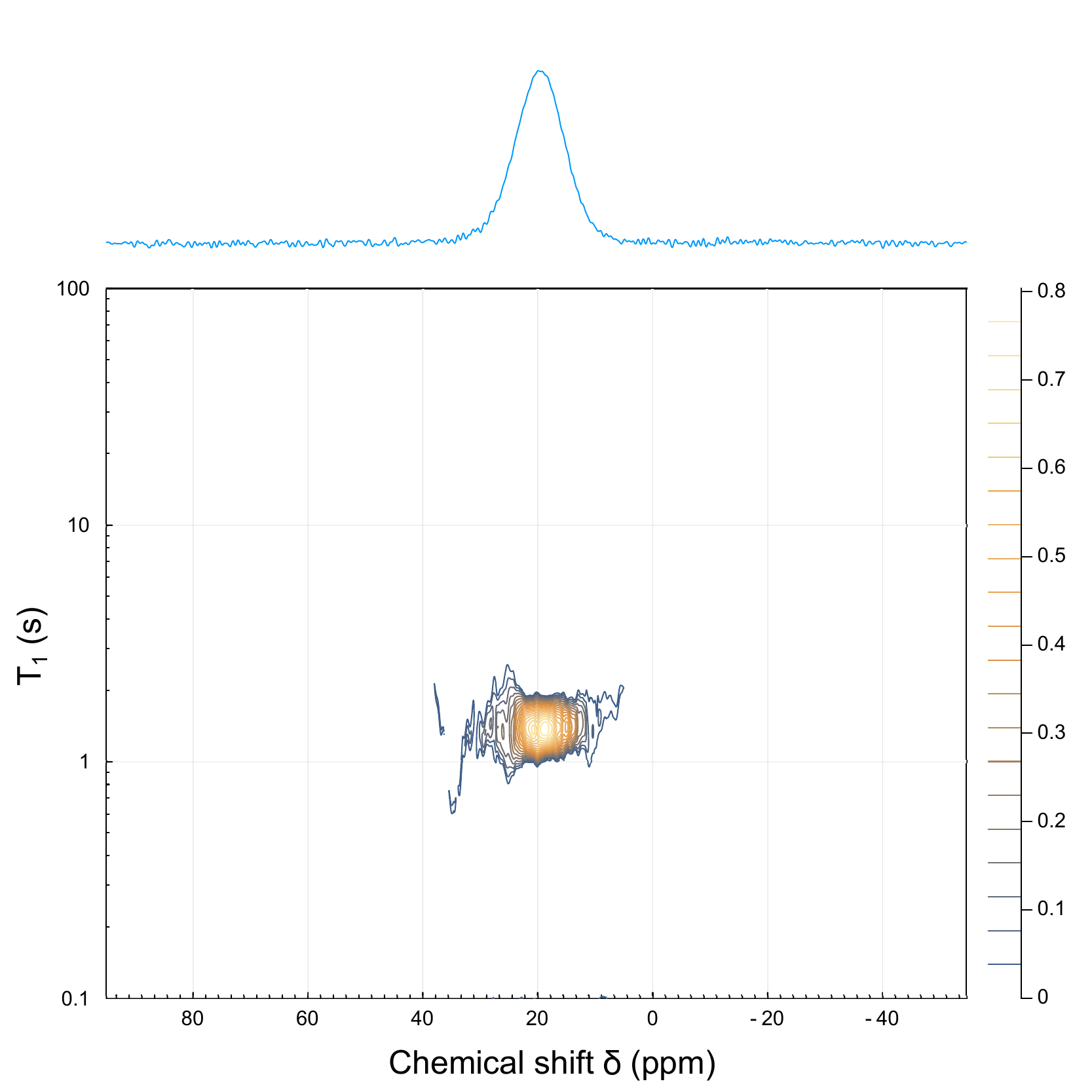}
    \caption{$\mathrm{^3He}$ RAS NMR spectrum~\cite{lupulescu_relaxation-assisted_2003} at 290~K. The 290~K NMR spectrum, extracted from the last slice of the inversion recovery data used as the input dataset, is displayed above the contour map. The relaxation time constant $T_1$ does not display a dependence on spectral frequency at this temperature.}
    \label{fig:ilft_290}
\end{figure}

\begin{figure}[ht]
    \centering
    \includegraphics[width=0.9\textwidth]{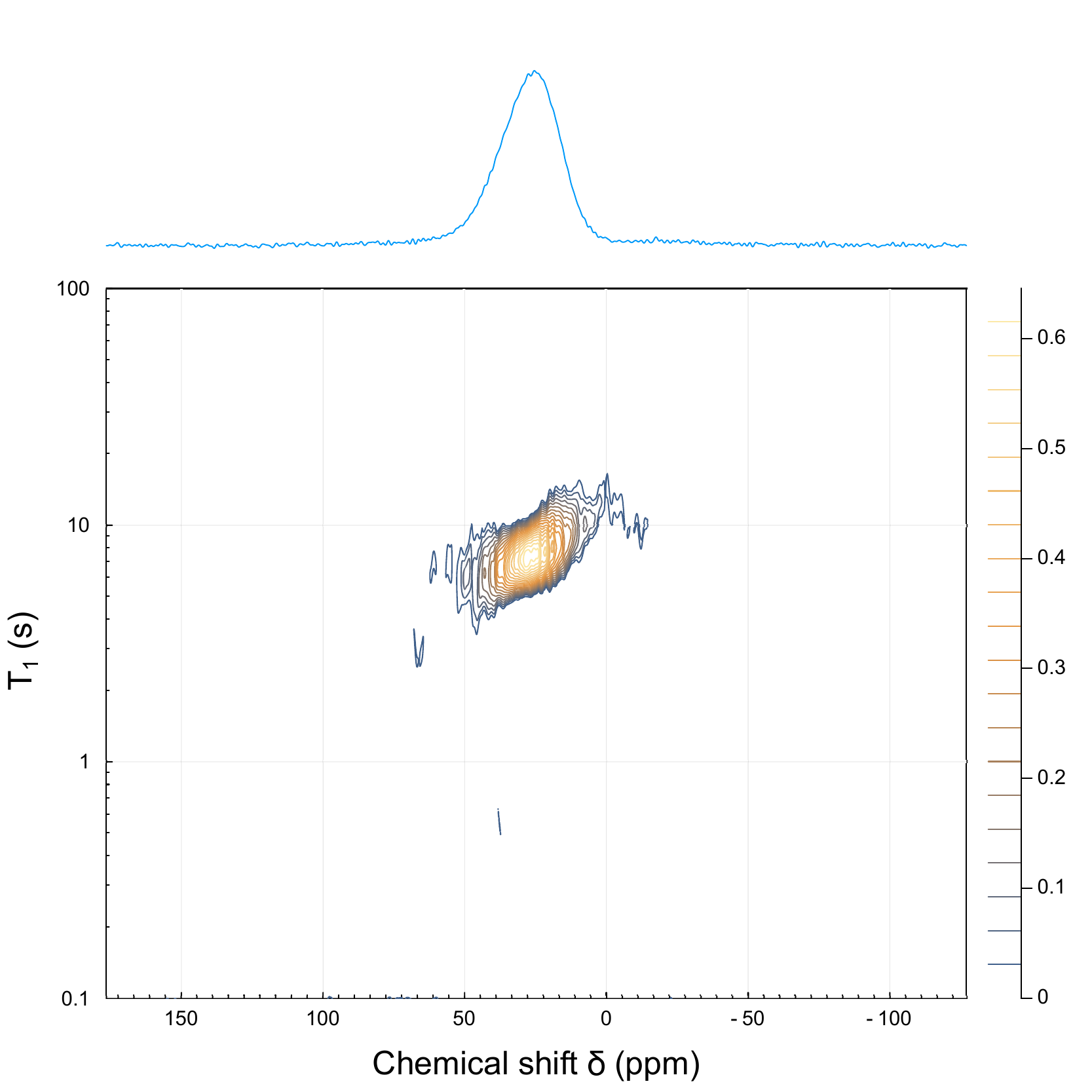}
    \caption{$\mathrm{^3He}$ RAS NMR spectrum~\cite{lupulescu_relaxation-assisted_2003} at 60~K. The 60~K NMR spectrum, extracted from the last slice of the inversion recovery data used as the input dataset, is displayed above the contour map. The skewed shape of the peak indicates that the relaxation time constant $T_1$ displays a significant dependence on spectral frequency, even at a temperature more than 30~K above $T_C$.}
    \label{fig:ilft_60}
\end{figure}

\hyperref[{fig:ilft_15}]{Figure~\ref*{fig:ilft_15}} shows that at 15~K the value of $T_1$  varies continously across the NMR spectrum from 30\,s at 53\,ppm to 110\,s at -29\,ppm. As the magnetic field distribution inside type-II superconductors at externally applied fields in the $H_{C1} < H_0 < H_{C2}$ range is highly inhomogeneous due to the vortex lattice \cite{brandt_Magneticfieldvariance_1991}, this frequency-dependent relaxation directly maps to a spatially dependent relaxation. Such inhomogeneous $T_1$ behaviour is commonly observed in cuprates and other d-wave superconductors~\cite{wortis_Spinlatticerelaxation_2000, mounce_Nuclearmagnetic_2011}, where it acts as a probe of vortex excitations. In s-wave superconductors, this is explained in literature as fast relaxation within the vortex cores that spreads to the bulk through spin diffusion, and has been observed in $\mathrm{^{31}P}$ NMR on skutterudites \cite{nakai_SiteselectiveNMR_2008}.

The spectral frequency-dependence of $T_1$ is not seen at 290~K (\hyperref[{fig:ilft_290}]{Figure~\ref*{fig:ilft_290}}), but begins to develop at 60~K, well above $T_C$ (\hyperref[{fig:ilft_60}]{Figure~\ref*{fig:ilft_60}}).
Indeed, the $T_1$ map at 60~K is qualitatively similar to that observed in the superconducting state at 15~K, albeit with a much smaller range of $T_1$ values. 

Pennington \textit{et al.}~\cite{pennington_13NMR_1996} reported a superficially similar variation in normal-state $\mathrm{^{13}C}$ $T_1$ across the spectrum, which they attributed to a dipolar contribution to the hyperfine interactions for $\mathrm{^{13}C}$, causing a correlated anisotropy in the spectral shift and the spin-lattice relaxation. However, in our case, the $T_1$ values are \emph{inversely} correlated with frequency, contrary to the results reported in ref.~\cite{pennington_13NMR_1996}. Furthermore, the dipole-dipole mechanism proposed in \cite{pennington_13NMR_1996} for the case of $\mathrm{^{13}C}$ appears to be unavailable for the $\mathrm{^3He}$ nuclei, which are located at the centres of the fulleride cages. 

The mechanism of the frequency-dependent $T_1$ is not known at the current time. Nevertheless, the detection of an unusual effect that is very strong below $T_C$ but which persists in an attenuated form over a wide temperature range above $T_C$ is potentially significant. The phenomenon is reminiscent of superconducting fluctuations at temperatures much higher than $T_C$, recently detected by transport measurements on $\mathrm{K_3C_{60}}$~\cite{jotzu_SuperconductingFluctuations_2023}. Determining whether these phenomena have a common origin is a matter for future research. 

\section{Conclusion}\label{conclusion}

A novel ``endofulleride'' material $\mathrm{Rb_3(^3He@C_{60})}$ was synthesised, and the structure and bulk magnetic properties of the parent fulleride were shown to be essentially unperturbed by the presence of 
the endohedral $\mathrm{^3He}$ atom. The $\mathrm{^3He}$ nucleus was used as an NMR probe in both the normal and the superconducting states of the material, and minor deviations from ideal~Knight shift and~Korringa behaviour were observed in the normal state, consistent with prior literature on $\mathrm{Rb_3C_{60}}$. Superconducting state measurements display features in the $\mathrm{^3He}$ NMR spectrum which are consistent with known features of vortex dynamics, such as the apparent co-existence of a vortex solid and vortex liquid phase. 
Similar behaviour has been observed by $\mathrm{^{13}C}$ and $\mathrm{^{87}Rb}$ relaxation and 2D exchange experiments in $\mathrm{Rb_3C_{60}}$ \cite{zimmer_Vortexdynamics_1996}. However, to our knowledge, this is the first time such phenomena have been directly observed in the NMR spectra of fulleride superconductors. The $T_1$ relaxation times in the superconducting state vary strongly across the NMR spectrum, a phenomenon that has been previously observed in d-wave and some s-wave superconductors but not in fullerides. Surprisingly, this was seen to persist even above $T_C$. While recent reports suggest the possibility of superconducting fluctuations above $T_C$ in $\mathrm{K_3C_{60}}$~\cite{jotzu_SuperconductingFluctuations_2023}, more detailed study is needed to conclusively determine the origin of the spectral distribution of $T_1$ values in $\mathrm{Rb_3(^3He@C_{60})}$.

Endohedral $\mathrm{^3He}$ has proven to be a valuable and sensitive addition to the existing NMR tools for studying the properties of fullerides. While $\mathrm{Rb_3C_{60}}$ is at the limit of conventionally understood superconductivity in fullerides, materials like $\mathrm{Cs_3C_{60}}$ and $\mathrm{Rb_xCs_{3-x}C_{60}}$ are known to exhibit more unusual properties including strong electron correlations. Introducing endohedral $\mathrm{^3He}$ into such materials raises interesting possibilities for their NMR characterisation.

\section*{Acknowledgments}
The authors are grateful to Dr.~Richard Bounds, Dr.~Karel~Kou{\v{r}}il, and Dr.~Roland Thoma for their prior work in the group on similar endofulleride systems, and to Dr.~Mark E. Light for assistance with the X-ray diffraction measurements. The authors would like to thank Prof.~Kosmas Prassides for valuable discussions and insight into alkali metal fulleride systems.

\section*{Declarations}

\subsection*{Funding}
This work was supported by the Engineering and Physical Sciences Research Council (UK), grant numbers EP/T004320/1, EP/P009980/1, EP/M001962/1, EP/K00509X/1 and EP/P030491/1. The powder diffraction facilities (Rigaku Smartlab) were supported through grant number EP/K009877/1.
\subsection*{Authors' contributions}
MS and GRB performed the NMR experiments and data analysis. MCW and GH synthesised the fullerene precursor and FG synthesised the fulleride material. FG performed the X-ray diffraction and MRL performed the magnetometry measurements. MHL, RJW and MC were involved in conceptualising and acquiring funding for the project and supervised the research. MS wrote the first draft. MS, GRB, MRL, RJW and MHL reviewed and edited the manuscript.


\end{document}


\title[Supplementary Information]{Supplementary Information}

\maketitle

\section{Variable Temperature NMR Probe}
\begin{figure}[ht]
    \centering
    \includegraphics[width=0.9\textwidth]{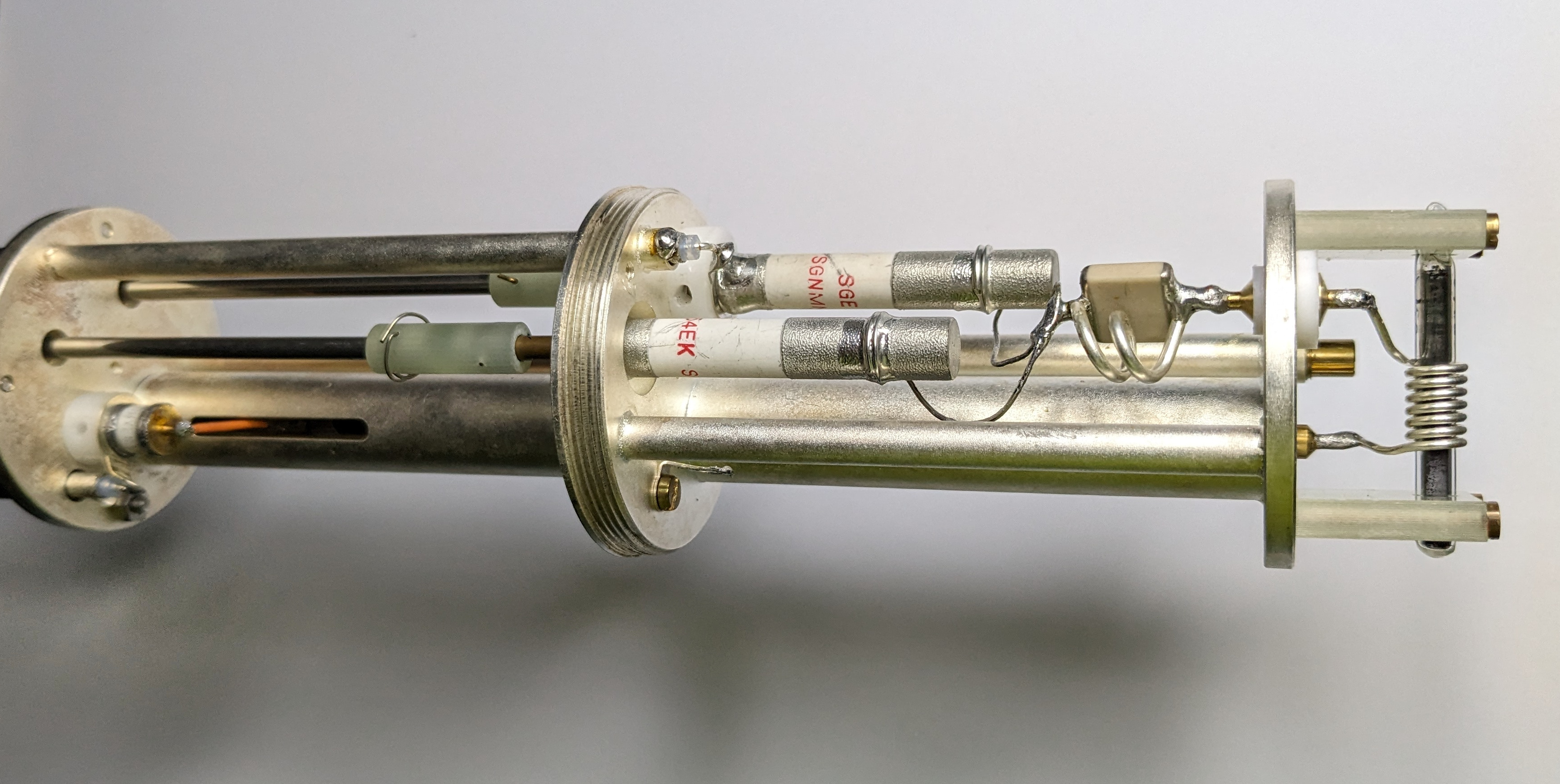}
    \caption{Photograph of the $\mathrm{^3He/X}$ NMR probehead with a sealed sample inserted into the RF coil.}
    \label{fig:probe_NMR}
\end{figure}

The NMR probe used for the cryogenic experiments is a dual channel $\mathrm{^3He/X}$ probe designed by NMR Service GmbH for a $\mathrm{^3He}$ resonance frequency of 457~MHz. The  probe is constructed from non-magnetic stainless steel and brass, and is fitted with a Cernox temperature sensor. The circuit has a single RF coil in the form of a solenoid, wound from silver-coated copper wire. The broadband channel uses two Sprague-Goodman trimmer capacitors while the $\mathrm{^3He}$ channel uses a Voltronics trimmer capacitor and a PTFE/air transmission line of variable length for the tuning and matching components. An additional high-frequency coil can be fitted to the broadband channel to increase its tuning range. A photograph of the probehead is presented in \hyperref[{fig:probe_NMR}]{Figure~\ref*{fig:probe_NMR}}.

\begin{figure}[ht]
    \centering
    \includegraphics[width=0.9\textwidth]{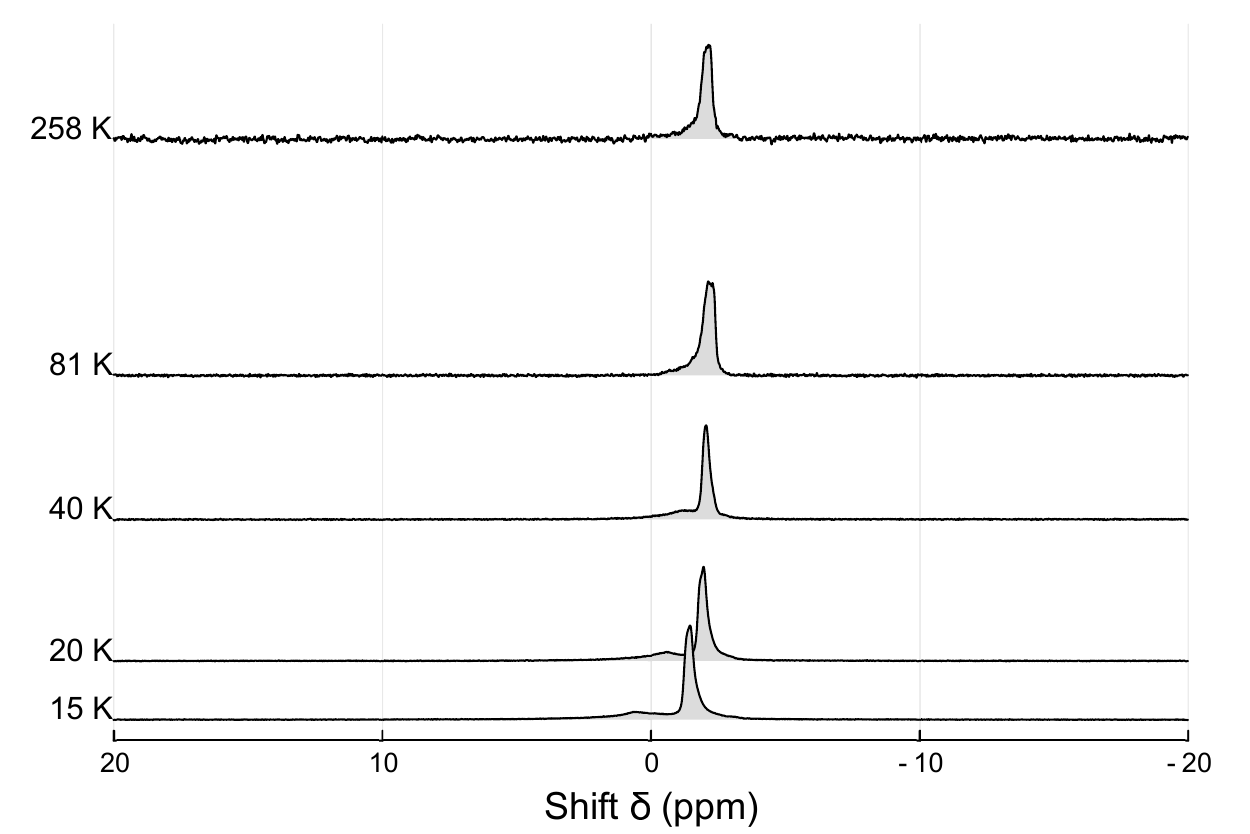}
    \caption{$\mathrm{^3He}$ NMR spectra at 14.1~T of $\mathrm{^3He}$ gas from 15~K to 258~K presented as a waterfall plot, acquired as an average of 8 scans with a 60~s recycling delay. The vertical scale of each spectrum has been adjusted to give equal peak heights across the series presented.}
    \label{fig:waterfall_gas}
\end{figure}

To test the performance of the NMR probe and cryogenic setup, reference $\mathrm{^3He}$ NMR spectra were acquired at various temperatures between 15 and 260~K. No attempt was made to re-shim between temperature points. The sample used was 0.25~mbar of $\mathrm{^3He}$ gas, flame sealed in a borosilicate tube of 4~mm outer diameter and 2~mm inner diameter. NMR spectra were acquired with 8 scans and a 60~s recycling delay and are shown in \hyperref[{fig:waterfall_gas}]{Figure~\ref*{fig:waterfall_gas}}. The integrated areas display the expected Boltzmann temperature dependence. The spectra in the figure have been normalised with respect to peak height for clarity. A relatively narrow peak of 0.4~ppm in linewidth is obtained. Below 40~K, some distortion in the lineshape is seen and a shoulder develops near the base of the spectrum. The width of this shoulder increases as the temperature is further decreased, and reaches about 2~ppm at 15~K. This lineshape distortion is negligible compared to the widths of the $\mathrm{Rb_3({}^3He@C_{60})}$ spectra obtained at the corresponding temperatures. Note that the shift of 0.6~ppm observed for the $\mathrm{^3He}$ gas signal between 15~K and 258~K is much less than the shift of 8~ppm observed for the $\mathrm{^3He}$ resonance in $\mathrm{Rb_3({}^3He@C_{60})}$ (Figure 3 of the main manuscript).

\section{X-ray Diffraction}

\begin{figure}[ht]
    \centering
    \includegraphics[width=0.9\textwidth]{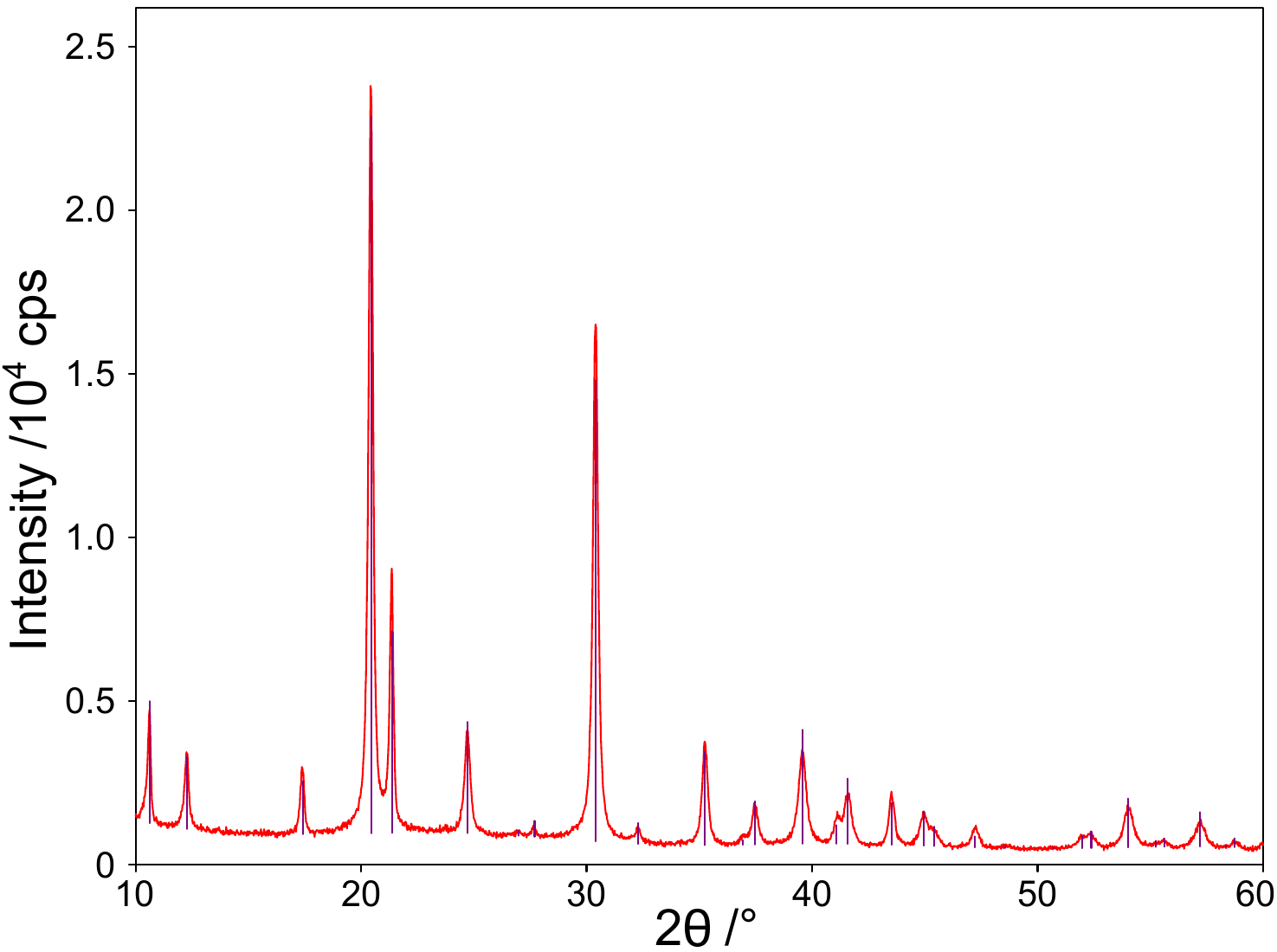}
    \caption{Powder X-ray diffraction pattern of the $\mathrm{Rb_3({}^3He@C_{60})}$ sample in red, along with expected peak positions of $\mathrm{Rb_3C_{60}}$ marked as black lines (extracted from a standard database).}
    \label{fig:XRDSpec}
\end{figure}

The crystallographic purity of the synthesised $\mathrm{Rb_3({}^3He@C_{60})}$ samples (with $22\%$ filling factor) was verified using powder X-ray diffraction. Measurements were carried out using a Rigaku X-ray diffractometer. Samples from the same synthesis batch as the NMR sample were packed in standard borosilicate capillaries and sealed with epoxy under inert atmosphere. Sample masses of $\sim2$~mg were used, and a representative pattern is shown in \hyperref[{fig:XRDSpec}]{Figure~\ref*{fig:XRDSpec}}. The measured pattern appears to match well with known literature~\cite{zhou_StructuresC60_1992} for $\mathrm{Rb_3C_{60}}$ with very few spurious peaks, indicating that the sample is of high purity and contains negligible amounts of residual $\mathrm{Rb_6C_{60}}$ and other $\mathrm{Rb_xC_{60}}$ impurities. The absence of an amorphous signature near $2\theta$ of $20\degree-25\degree$ indicates the very high crystallinity of the sample. Since no significant difference was observed between the material containing endohedral $\mathrm{^3He}$ and empty-cage material, the patterns were matched using Rigaku's PDXL-2 software against data for empty-cage $\mathrm{Rb_3C_{60}}$ from a standard database.

\section{Rubidium-87 NMR}
\begin{figure}[ht]
    \centering
    \includegraphics[width=0.9\textwidth]{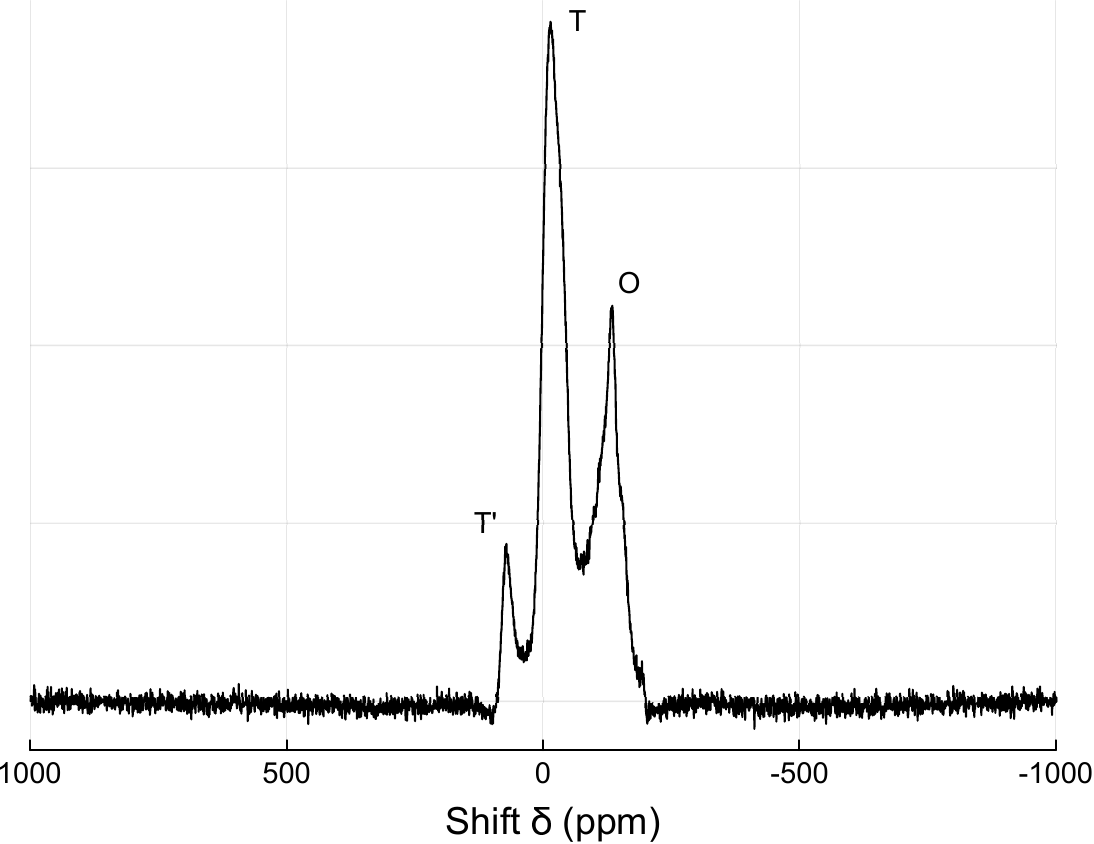}
    \caption{$\mathrm{^{87}Rb}$ NMR spectrum of $\mathrm{Rb_3({}^3He@C_{60})}$ at 14.1~T and 298~K, acquired as a Hahn echo with full echo delay $2\tau$ of 200~$\mathrm{\mu s}$. The three peaks labelled T, T' and O correspond to the two tetrahedral sites and one octahedral site occupied by the Rb nuclei.}
    \label{fig:spectrum_Rb}
\end{figure}

A $\mathrm{^{87}Rb}$ NMR spectrum of the $\mathrm{Rb_3({}^3He@C_{60})}$ sample was acquired at 298~K using a Hahn echo sequence with a nutation frequency of 50~kHz and a full echo delay $2\tau$ of 200~$\mathrm{\mu s}$, and is presented in \hyperref[{fig:spectrum_Rb}]{Figure~\ref*{fig:spectrum_Rb}}. The central transitions of the three sites known in $\mathrm{Rb_3C_{60}}$ \cite{zimmer_Analysis87_1993}, two tetrahedral (T, T') and one octahedral (O), are marked on the spectrum. No additional features from other compositions of $\mathrm{Rb_xC_{60}}$ with x not equal to 3 are observed. The observed $\mathrm{^{87}Rb}$ NMR spectrum is in good agreement with previous literature report for $\mathrm{Rb_3C_{60}}$ \cite{zimmer_Analysis87_1993}.

\section{Magnetometry}

\begin{figure}[ht]
    \centering
    \includegraphics[width=0.8\textwidth]{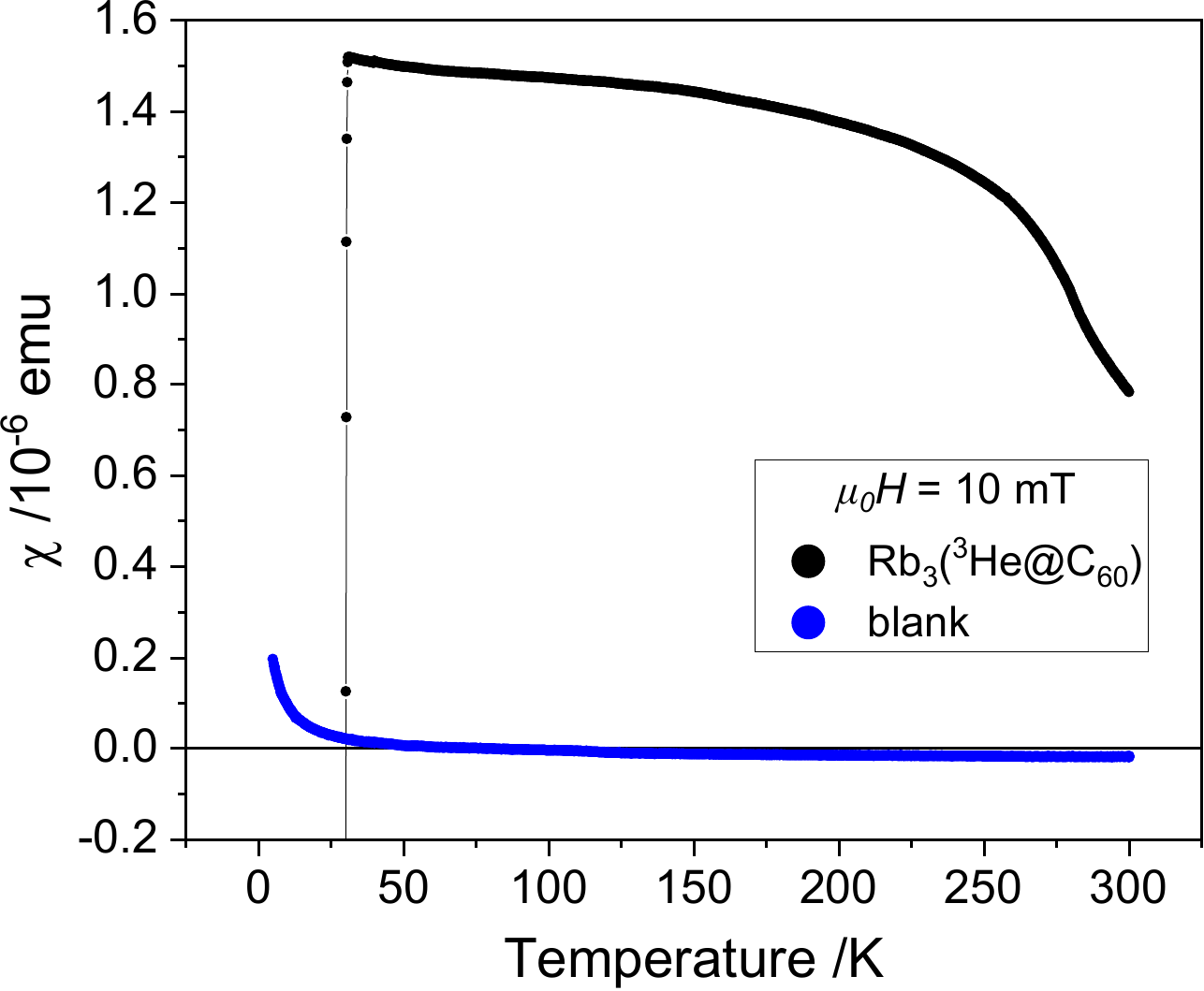}
    \caption{Susceptibility $\chi$ versus temperature of the $\mathrm{Rb_3({}^3He@C_{60})}$ sample, in black, and of an empty tube of the same geometry filled with 0.25~bar of $\mathrm{^3He}$ gas, in blue, at an applied magnetic field $\mu_0H$ of 10~mT. The susceptibility is presented in unnormalised units of emu as the mass and density of paramagnetic centres in the glass are unknown.}
    \label{fig:blank}
\end{figure}

To ensure that the sample susceptibility as measured by MPMS is not significantly affected by contributions from the glass sample holder, a reference measurement was performed on a blank sample holder of the same geometry as the sample, sealed under 0.25~bar of $\mathrm{{}^3He}$ gas. The resulting susceptibility is presented in \hyperref[{fig:blank}]{Figure~\ref*{fig:blank}} along with the sample susceptibility measured at the same applied field $\mu_0H$ of 10~mT. While the glass has some measurable susceptibility, this is rather small when compared to the sample the normal state, and negligible when compared to the superconducting state.

\section{RAS NMR}

$\mathrm{^3He}$ relaxation-assisted separation (RAS) NMR spectra of $\mathrm{Rb_3({}^3He@C_{60})}$, as described in section 3.3 of the main text, are presented at 30~K, 35~K and 40~K below. The $T_1$ is seen to depend on spectral position, to a similar extent as seen at 60~K in figure 7 of the main text.

\begin{figure}
    \centering
    \includegraphics[width=0.9\textwidth]{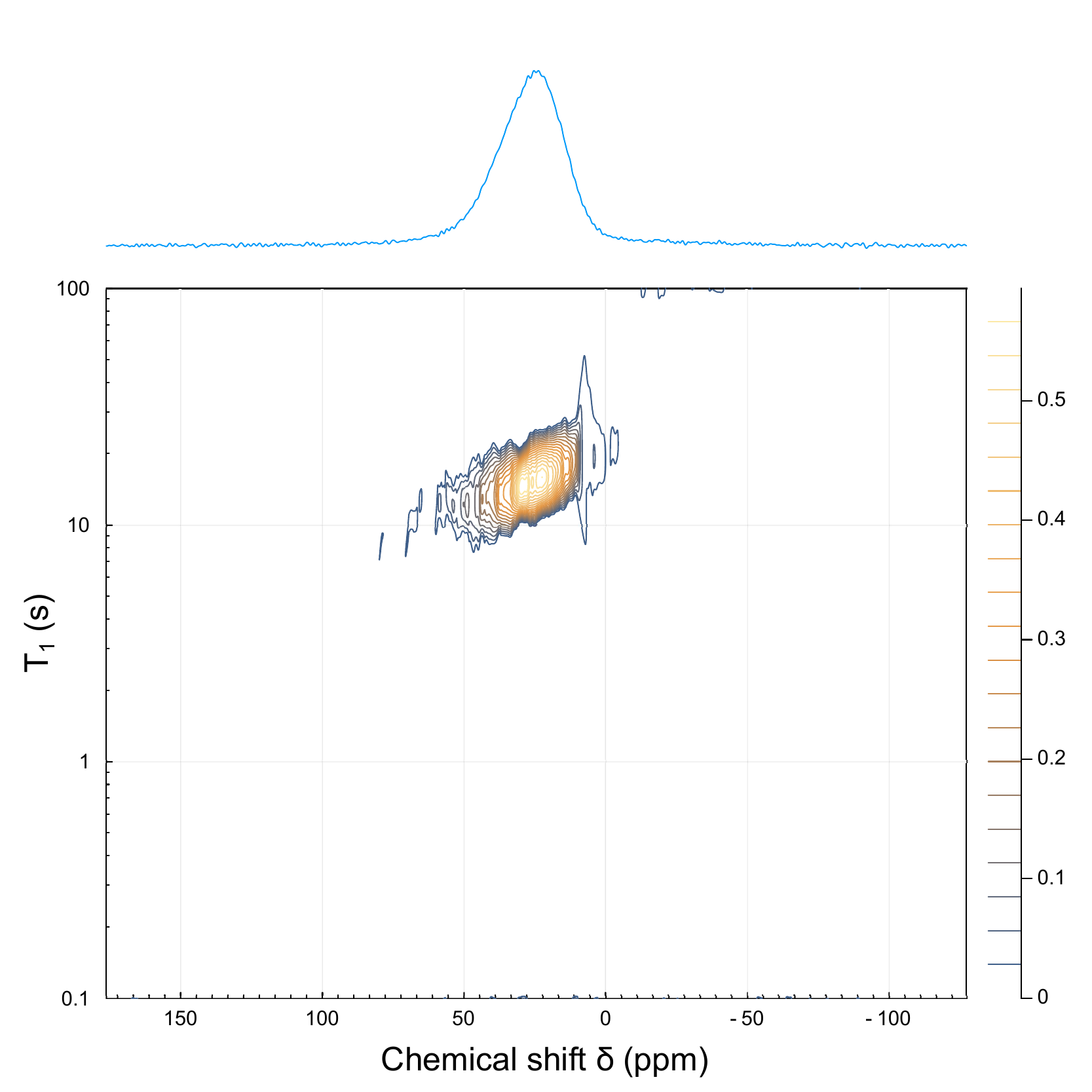}
    \caption{$\mathrm{^3He}$ RAS NMR spectrum at 30~K. The 30~K NMR spectrum, extracted from the last slice of the inversion recovery data used as the input dataset, is displayed above the contour map.}
    \label{fig:ILFT_30}
\end{figure}

\begin{figure}
    \centering
    \includegraphics[width=0.9\textwidth]{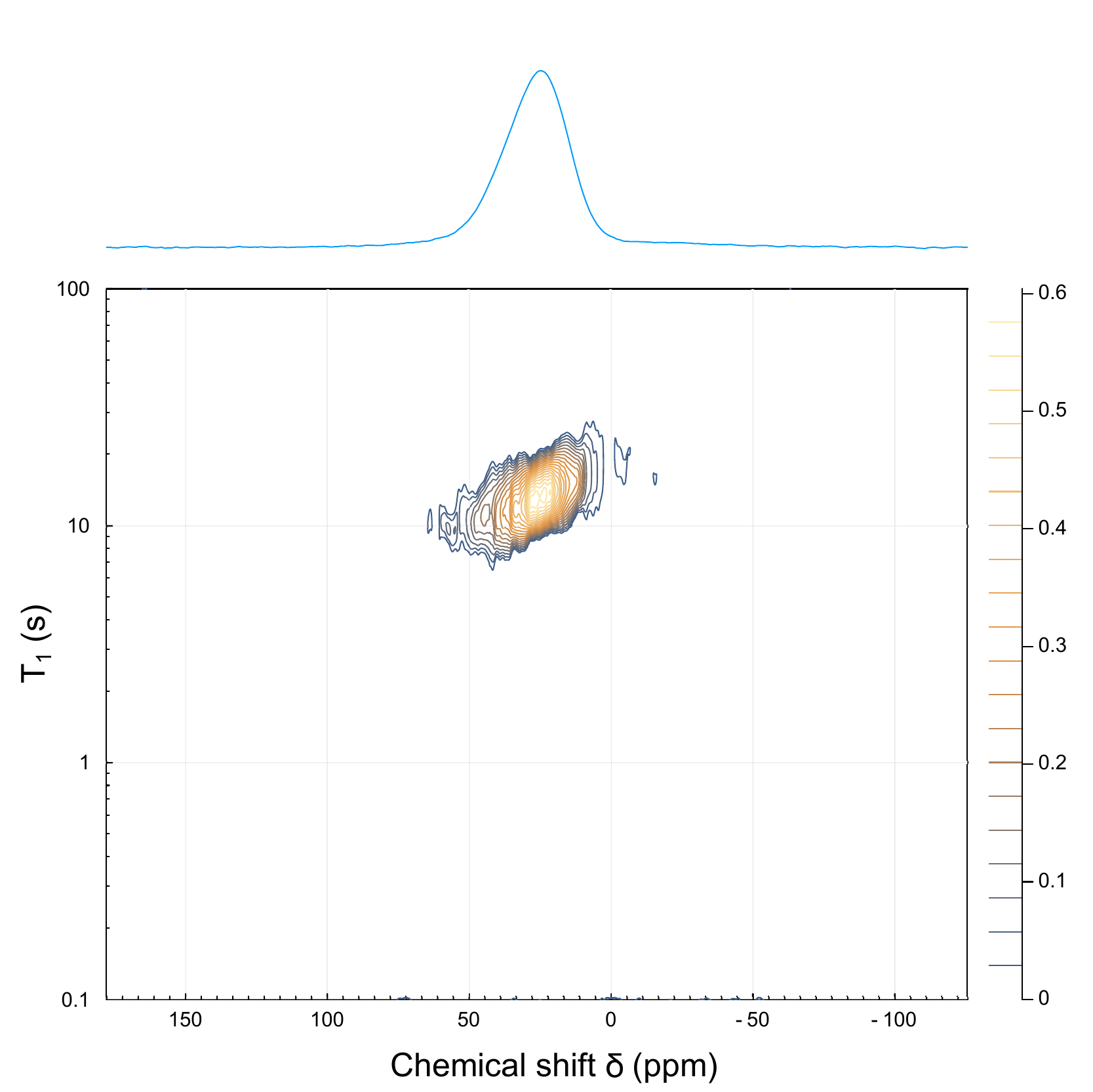}
    \caption{$\mathrm{^3He}$ RAS NMR spectrum at 35~K. The 35~K NMR spectrum, extracted from the last slice of the inversion recovery data used as the input dataset, is displayed above the contour map.}
    \label{fig:ILFT_35}
\end{figure}

\begin{figure}
    \centering
    \includegraphics[width=0.9\textwidth]{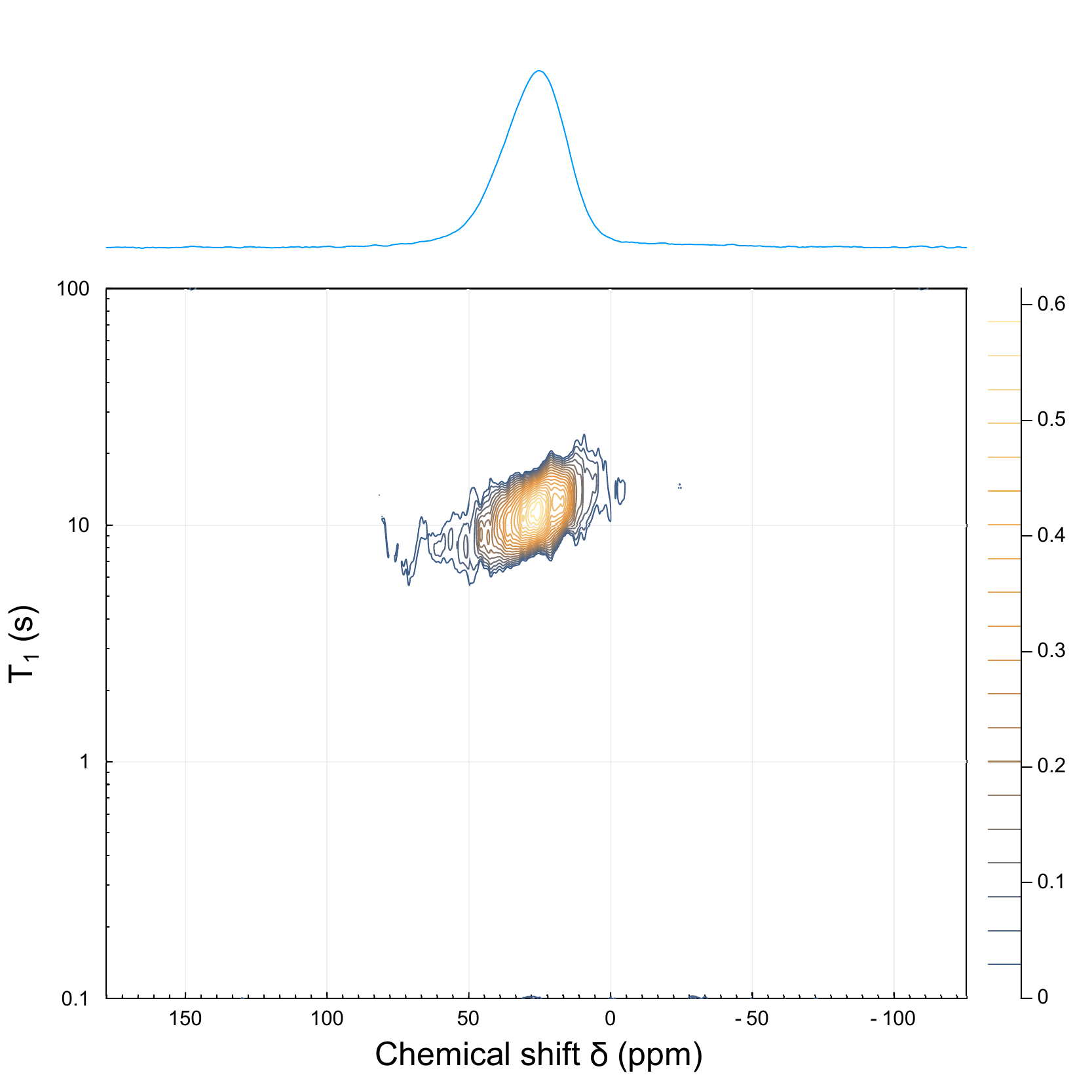}
    \caption{$\mathrm{^3He}$ RAS NMR spectrum at 40~K. The 40~K NMR spectrum, extracted from the last slice of the inversion recovery data used as the input dataset, is displayed above the contour map.}
    \label{fig:ILFT_40}
\end{figure}
